\documentstyle[eqsecnum,preprint,aps,epsfig]{revtex}
\tightenlines

\newcommand{\beq}{\begin{equation}}
\newcommand{\eeq}{\end{equation}}
\newcommand{\beqa}{\begin{eqnarray}}
\newcommand{\eeqa}{\end{eqnarray}}
\def\nn{\nonumber}
\def\l({\left(}
\def\r){\right)}

\title{Fermion masses and quantum numbers from extra dimensions.}
\author{Andrey Neronov}
\address{Theoretische Physik, Universit\"at M\"unchen, Theresienstr., 37, 
80333, Munich, Germany} 
\begin{document}
\maketitle
\draft

\narrowtext

\begin{abstract}
We study the localization of  fermions on a brane embedded in a 
space-time with $AdS_n\times M^k$ geometry.
 Quantum numbers of localized fermions are associated with 
their rotation momenta around the brane. Fermions with different quantum 
numbers have different higher-dimensional profiles. 
Fermion masses and mixings, which are proportional to the overlap 
of higher-dimensional profiles of the fermions, depend on the fermion 
quantum numbers. 
\end{abstract}

\section{introduction.}

The problem of explaining the hierarchy of fermion masses and mixings is one 
of the longstanding problems of particle physics which can not be resolved 
within the standard model  \cite{mixings}. This  problem 
 becomes especially interesting in the light of recent results on nonzero 
neutrino masses \cite{neutrino}.
There exist different approaches 
to explanation of the hierarchy of fermion masses, 
like models with  additional ``horizontal'' 
symmetry \cite{generation}, or models where the hierarchy 
is generated by the ``seesaw'' mechanism \cite{seesaw}, the last class of 
models is particularly popular for explanation of smallness of neutrino masses.
Apart from the seesaw mechanism the smallness of neutrino masses can be related to 
the possible existence of large extra dimensions \cite{sterilebulk}. The 
idea is to assume  that the right-handed 
sterile neutrino can propagate in higher-dimensional bulk while the 
left-handed neutrinos are  confined to a four-dimensional brane so that 
the higher-dimensional profiles 
of left and right neutrinos have small overlap. 
Actually, it is possible to relate the whole hierarchy of fermion masses 
and mixings to the differences in the overlaps of higher-dimensional
fermion profiles  \cite{schmaltz,dva}.

What can be the reason for different fermions to have different 
higher-dimensional profiles? What mechanism can be responsible for the fact 
that a fermion which is neutral with respect to the  
$SU(3)\times SU(2)\times U(1)$ group of the standard 
model is not localized on the brane? 
   
If a brane is embedded as a surface in a higher-dimensional space, 
matter fields bound to the brane are naturally classified by the values of 
 their momenta of rotation around the brane. Indeed, 
in quite general settings, the space-time around the brane possess a 
rotation  symmetry. For example, in the case of just two extra dimensions
the metric around the brane has the form  
\beq
\label{metric}
ds^2=e^{\nu(\rho)}\eta_{\mu\nu}dx^\mu dx^\nu+e^{\lambda(\rho)}
d\rho^2+e^{\mu(\rho)}dy^2
\eeq
where $\eta_{\mu\nu}=\mbox{diag}(-1, 1, 1, 1)$ is the four-dimensional Minkowsky metric and functions $\nu(\rho), \lambda(\rho), \mu(\rho)$ are determined 
from 
Einstein equations. The coordinate $\rho$ measures the distance from the 
brane placed at $\rho=0$ and the coordinate $y$ is periodic $y\in [0, 2\pi R_y)$. The metric 
(\ref{metric}) possesses a $U(1)_y$ 
symmetry of rotations in $y$ direction
\beq
\label{yy}
y\rightarrow y+const
\eeq
As in usual quantum mechanics, matter fields $\Psi$ bound to the brane  
are characterized by 
different values $q$ of the rotation momentum in the $y$ direction
\beq
\Psi_q(\rho, y)=F(\rho)e^{iqy}
\eeq 
From  the four-dimensional point of view, this rotation momentum is a quantum 
number of observable four-dimensional particles.
The profiles $\Psi_q(\rho, y)$ 
(``wave functions'') of the bound states 
depend on the values of the rotation momentum. This means that 
particles with different quantum numbers $q$ have different 
higher-dimensional profiles \cite{neronov}.

In what follows we consider 
branes embedded into higher-dimensional spaces with (asymptotic) 
$AdS_n\times M^k$ 
geometry of direct product of 
 $n$-dimensional anti-deSitter space and compact manifold $M^k$. Such 
spaces
arise naturally in Freund-Rubin compactifications \cite{freund} of 
higher-dimensional supergravities \cite{sugra} 
and received a considerable attention 
recently after the discovery of $AdS/$CFT correspondence \cite{adscft}  
and observation that gravity can be localized on a brane in 
five-dimensional  \cite{rs} or higher-dimensional \cite{6dads} anti-deSitter
space.

In section \ref{ads6} we show that in six-dimensional space (\ref{metric}) 
fermion zero modes which are charged with respect to $U(1)_y$ group 
of rotations in $y$ direction (\ref{yy}) can be localized on the brane, 
while the state neutral with respect to $U(1)_y$ propagates in the bulk. 
 In Section \ref{masses} we discuss the breaking 
of rotation symmetry $U(1)_y$ (\ref{yy}) and derive a formula relating 
the effective four-dimensional mass of a localized fermion state
 and charges $q_L$ and $q_R$ of its left and right handed components. We also 
consider the case when the brane  is 
embedded in seven-dimensional  space with $AdS_7$ geometry. In this case 
localized fermions carry $U(1)\times U(1)$ charges. The extra 
$U(1)$ symmetry can be associated with the  ``horizontal'' symmetry 
\cite{generation} needed to distinguish between fermions from different 
generations. We show that in this model the masses of localized fermions are
arranged hierarchically.
In Section \ref{monop} 
we consider a more complicated model where the localized fermions carry
$SU(2)\times U(1)$ quantum numbers. In this case the brane is embedded into 
space-time which becomes the direct product of $AdS$ space with a 
two-dimensional sphere $S^2$ far away from the brane. 
We find the higher-dimensional  profiles of 
$SU(2)$ singlet and doublet states localized on the brane. If the rotation 
symmetry $SU(2)$ is broken, effective four-dimensional mixing between 
singlet and doublet states becomes nonzero and one of the doublet components 
becomes massive. The other component (``left-handed neutrino'') 
is not mixed with the states localized on the brane and, 
therefore, its  mass is naturally small.

\section{Fermion quantum numbers as momenta of rotation around the brane.}
\label{ads6}

Let us consider a brane embedded in the 
space-time  (\ref{metric}). Suppose that all the matter fields are localized 
in a region   
$0\le\rho\le\rho_0$ while the metric outside the 
brane, $\rho_0<\rho<\infty$, is a solution 
of vacuum Einstein equations, possibly with a cosmological constant term. 
In the most simple case the bulk metric is isometric to the metric of 
the six-dimensional anti-deSitter space $AdS_6$ 
\beq
\label{ads66}
ds^2=\frac{1}{(\kappa \rho)^2}\l(\eta_{\mu\nu}dx^\mu dx^\nu+d\rho^2+dy^2\r)
\eeq
where $\kappa$ is the inverse curvature radius of anti-deSitter space. 

The massless modes of higher-dimensional Dirac field can be naturally 
localized on the brane in a space-time (\ref{metric}) with nontrivial warp 
factors $\nu(\rho),\lambda(\rho),\mu(\rho)$ \cite{neronov}. 
In order to see this let us consider the 
higher-dimensional Dirac equation
\begin{equation}
\label{dirac}
\Gamma^A D_A\Psi=0
\end{equation}
The six-dimensional gamma matrices $\Gamma^A$ are defined 
with the help of the vielbein $E^A_{\hat B}$ and flat space gamma matrices 
$\Gamma^{\hat A}$
\begin{equation}
\Gamma^A=E^A_{\hat B}\Gamma^{\hat B}
\end{equation}
(the indexes with a hat are six-dimensional Lorentz indexes).
 The covariant derivative is defined as 
\begin{equation}
\label{covar}
D_A\Psi=\Psi_{,A}+\frac{1}{2}\omega^{\hat B\hat C}_A\sigma_{\hat B\hat C}\Psi
\end{equation}
where $\omega_A^{\hat B\hat C}$ is the spin connection expressed through the 
vielbein $E^A_{\hat B}$ and
$\sigma_{\hat B\hat C}=\frac{1}{4}\left[\Gamma_{\hat B}\Gamma_{\hat C}\right]$ 
are generators of the six-dimensional Lorentz group.
Taking the coordinate vielbein for the metric (\ref{metric})
\begin{eqnarray}
\label{viel}
E_\mu^{\hat \alpha}&=&e^{\nu/2}\delta^{\hat \alpha}_\mu\nonumber\\
E^{\hat \rho}_\rho&=&e^{\lambda/2}\\
E^{\hat y}_{y}&=&e^{\mu/2}
\end{eqnarray}
we find
\beq
\frac{1}{2}\omega_A^{\hat B\hat C}\Gamma^{A}\sigma_{\hat B\hat C}=
e^{-\lambda/2}\l(\nu'+\frac{\mu'}{4}\r)
\Gamma^{\hat r}
\eeq 
where the prime denotes the $\rho$ derivative.
Substituting the covariant derivative (\ref{covar}) into the Dirac equation 
(\ref{dirac}) we find 
\beq
\label{dirac1}
e^{-\nu/2}\Gamma^{\hat\mu}\Psi_{,\mu}+e^{-\lambda/2}
\Gamma^{\hat r}\l(\Psi'+\l(\nu'+\frac{\mu'}{4}\r)
\Psi\r)+e^{-\mu/2}\Gamma^{\hat y}\Psi_{,y}=0
\eeq
The six-dimensional spinor $\Psi$ can be decomposed on a four-component
spinor $\psi$ and two-component spinor $\tilde \Psi$:
\beq
\label{splitting}
\Psi=\psi(x)\otimes \tilde\Psi(r, y)
\eeq
If we are interested in the massless solutions of Dirac equation, 
the four-dimensional spinor $\psi$ satisfies 
\beq
\Gamma^{\hat\mu}\psi_{,\mu}=0
\eeq
and the equation (\ref{dirac1}) reduces to
\beq
\Gamma^{\hat r}\l(\tilde \Psi'+\l(\nu'+\frac{\mu'}{4}\r)
\tilde \Psi\r)+e^{(\lambda-\mu)/2}\Gamma^{\hat y}\tilde \Psi_{,y}=0
\eeq
We can expand the solutions over the states with fixed rotation 
momentum $q_y$ in the $y$ direction
\beq
\label{prof}
\tilde \Psi=e^{-\nu-\mu/4}\exp\left\{\frac{iq_yy}{R_y}\right\}
F(\rho)
\eeq
($F$ is a two-component spinor).
If we take the two gamma matrices $\Gamma^{\hat r}$ and  $\Gamma^{\hat y}$ 
to be
\beq
\Gamma^{\hat r}=\l(\begin{array}{cc}0&1\\1&0\end{array}\r); \ \ 
\Gamma^{\hat y}=\l(\begin{array}{cc}0&-i\\i&0\end{array}\r)
\eeq
we find that the two-component spinor $F=(f, g)$ satisfies the equations
\beqa
f'-e^{(\lambda-\mu)/2}\frac{q_y}{R_y}f&=&0\nn\\
g'+e^{(\lambda-\mu)/2}\frac{q_y}{R_y}g&=&0
\eeqa
Thus, if $q_y>0$ and the bulk metric is isometric to anti-deSitter metric 
(\ref{ads66}), the spinor 
\beq
\label{profile}
\Psi_q=\sqrt{\frac{q_y}{\pi}}\frac{(\kappa \rho)^{5/2}}{R_y}
\l(\begin{array}{c}\displaystyle
0\\ \displaystyle\exp\left\{\frac{iq_yy}{R_y}-\frac{q(\rho-\rho_0)}{R_y}
\right\}\end{array}\r)\otimes \psi(x) 
\eeq
describes  a fermion state bound to the brane
(We postpone the discussion of the behavior of solutions of Dirac equation 
in the core of the brane $0\le \rho\le \rho_0$ till Section \ref{monop}). 
We have normalized the solution
(\ref{profile}) with respect to the scalar product
\beq
\label{norm}
\left<\Psi_q,\Psi_q\right>=\int d^4xd\rho dy\sqrt{-g}\ 
\overline\Psi_q\Gamma^0\Psi_q
\eeq

The rotation momentum $q_y$ is, in fact, a charge of the localized fermion 
mode 
with respect to Kaluza-Klein gauge field $A_\mu$ which corresponds to the 
$U(1)_y$ symmetry (\ref{yy}) of the metric (\ref{metric}). Indeed, the metric 
(\ref{metric}) possesses
a Killing vector
\beq
K=\partial_y
\eeq
which, according to  Kaluza-Klein mechanism leads to the existence of 
a gauge field $A_\mu$ in effective four-dimensional theory. This 
gauge field arises as a nondiagonal component
\beq
ds^2=e^{\nu}\eta_{\mu\nu}dx^\mu dx^\nu+e^{\lambda}d\rho^2+
e^{\mu}(dy+A_\mu dx^\mu)^2
\eeq 
 of higher-dimensional gravitational perturbations. The zero mode of the 
field $A_\mu$ can be localized on the brane \cite{neronov}. In this case 
$A_\mu$ can correspond to an observable $U(1)$ gauge field and 
the charge $q_y$   can be related to 
an observable quantum number of standard model fermions. Note that the 
$U(1)_y$--neutral state  with $q_y=0$ 
(``sterile neutrino'')
is not localized on the brane.

\section{Breaking of $U(1)$ rotation symmetry and generation of fermion masses.}
\label{masses}

We are interested in mixings  between differently charged fermions 
$\Psi_{q_1}$ and  $\Psi_{q_2}$ 
\beq
\label{mix}
S_{mix}\sim f \int d^4x d\rho dy \sqrt{-g}\ \overline \Psi_{q_1}
{\cal O}_{q_1q_2}\Psi_{q_2}
\eeq
here ${\cal O}_{q_1q_2}$ is a matrix with indexes which run through all  
fermion species (in our example through all possible $q_y$) and 
$f$ is a constant. Let us consider (\ref{mix}) in more details.
Higher-dimensional profiles (\ref{profile}) 
of the localized fermion zero modes $\Psi_q$ depend on their charges 
$q_y$.
The mixing (\ref{mix}) between modes with charges $q_1$ and $q_2$ 
is proportional to the 
integral over the extra dimensions of the overlap of the profiles 
of $\Psi_{q_1}$ and $\Psi_{q_2}$. This integral, in turn 
includes the integral over the circle $S^1$ parameterized by the coordinate 
$y$. Substituting the profiles (\ref{prof}) into (\ref{mix}) we find
\beq
\label{mix1}
S_{mix}\sim f\int d\rho\int\limits_0^{2\pi R_y} dy
\sqrt{-g}e^{i(q_2-q_1)y/R_y}\overline F_{q_1}(r)F_{q_2}(r)
\eeq 
It is not difficult to see that this integral vanishes if $q_1\not= q_2$ 
since the modes with different $q_y$ are orthogonal to each other.
Thus, when the symmetry of rotations around the brane is not broken,
the mixings between the modes with different charges vanish.

Suppose, that the symmetry of rotations $y\rightarrow y+const$ is  
broken by some mechanism.  We do not discuss here different possibilities 
for particular  mechanism of symmetry breaking in the context of 
theories with extra dimensions (see, for example \cite{breaking}). 
The symmetry 
breaking  results in appearance of a (fundamental or  effective) Higgs 
field which has nonzero $U(1)_y$ charge $p$
\beq
\label{higgs}
H_p=H_p(\rho)\exp\left\{\frac{ipy}{R_y}\right\}
\eeq
The Higgs field is coupled to the higher-dimensional Dirac field 
\beq
\label{mix2}
S_{mix}\sim f \int d^4 x d\rho dy \sqrt{-g}
H_p\overline\Psi_{q_1}\Psi_{q_2}
\eeq 
Substituting the profiles (\ref{profile}), (\ref{higgs}) into (\ref{mix2})
we find that the mixing between the modes $\Psi_{q_1}$, $\Psi_{q_2}$ with the charges $q_1, q_2$ such that
\beq
p-q_1+q_2=0
\eeq
does not vanish. 

Using the decomposition of the fermion field (\ref{splitting})  
we can write the mixing (\ref{mix2}) in the form
\beq
S_{mix}\sim M_{q_1q_2}\int d^4x \overline \psi_{q_1}(x)\psi_{q_2}(x)
\eeq
with
\beq
\label{mq}
M_{q_1q_2}=f \int d\rho dy \sqrt{-g}H_p\tilde{\overline \Psi}_{q_1}\tilde \Psi_{q_2}
\eeq
The mixing $M_{q_1q_2}$ 
can be naturally small, if the higher-dimensional  profiles of  
fermions $\Psi_{q_1},\Psi_{q_2}$ have small overlap with each other, 
or with the Higgs profile $H_p$.

The Dirac mass of a fermion is the mixing between its left and right handed 
components
$\Psi_L$ and $\Psi_R$. If the charges of $\Psi_L, \Psi_R$ are respectively
$q_L,$ and $q_R=q_L+p$,  we get from (\ref{mq})
\beq
\label{MD}
 M_D=f \int d\rho dy e^{2\nu+\lambda/2+\mu/2} 
H_p\tilde{\overline \Psi}_{q_L}\tilde \Psi_{q_R}
\eeq
Since we do not discuss the details of the symmetry breaking, we can not 
calculate the Higgs profile $H_p(\rho)$.
Let us consider two extreme possibilities. 
If the profile $H_p(\rho)$ is ``smooth'', 
\beq
H_p(\rho)\approx h=const
\eeq
we get, substituting  the profiles (\ref{profile}) into 
(\ref{MD}) and performing the integral
\beq
\label{m}
M_D= \frac{2f h \sqrt{q_Lq_R}}{\kappa R_y}e^{
(q_L+q_R)\rho_0/ R_y}\Gamma\l(0, \frac{(q_L+q_R)\rho_0}{R_y}\r) 
\eeq
where $\Gamma(0, x)$ is the incomplete Gamma-function. (We have supposed that 
the space-time metric outside the brane is the anti-deSitter metric 
(\ref{ads66}). $\rho_0$ is the thickness of the brane core.)
If the brane thickness 
$\rho_0$ is much larger than $R_y$ we get an approximate expression
\beq
\label{m1}
M_D\approx \frac{2 f h}{\kappa\rho_0}\frac{\sqrt{q_Lq_R}}{(q_L+q_R)}
\eeq 
If the profile $H_p(\rho)$ 
of the Higgs field is ``sharp'', that is peaked at a distance 
$\rho_h$ from the center of the brane 
\beq
\label{sharp}
H_p(\rho)\approx h\rho_h \delta(\rho-\rho_h)
\eeq
we get from (\ref{MD})
\beq
\label{m2}
M_D=\frac{2 fh\sqrt{q_Lq_R}}{\kappa R_y}
e^{-(q_L+q_R)(\rho_h-\rho_0)/R_y}
\eeq
In this case, if $\rho_h-\rho_0= (several) R_y$, the  masses of 
particles with different $q_L, q_R$ are arranged 
hierarchically.
The fact that the sharp higher-dimensional profile of the Higgs field 
leads to a hierarchical structure of four-dimensional fermion masses 
was noted in \cite{dva,troitsky}.

As a simple generalization of the above model let us consider  
a brane embedded in a seven-dimensional space with
$AdS_7$ geometry
\beq
\label{metric4}
ds^2=\frac{1}{(\kappa \rho)^2}\l(\eta_{\mu\nu}dx^\mu dx^\nu +dy^2+dz^2+
d\rho^2\r)
\eeq
The seven-dimensional anti-deSitter space 
can be obtained after compactification 
of eleven-dimensional supergravity \cite{freund}. Suppose that, as in the 
above example (\ref{metric}) the coordinate $\rho$ counts the distance 
from the brane and the coordinates $y$ and $z$ are periodic
with the periods 
$2\pi R_y, 2\pi R_z$. The symmetry group of rotations around the brane is
now $U(1)_y\times U(1)_z$ -- the product 
of rotations in $y$ and $z$ directions. 
The fermions bound to the brane possess two quantum numbers 
$q_y, q_z$ which are rotation momenta in $y$ and $z$ directions
\beq
\Psi=\exp\left\{\frac{iq_yy}{R_y}+\frac{iq_zz}{R_z}\right\}F(r)\otimes\psi(x)
\eeq
The Dirac equation for zero modes $\Gamma^{\hat\mu}\psi_{,\mu}=0$ is
reduced to
\beq
\Gamma^{\hat r}\l(F'-\frac{5}{\rho}
F\r)+\frac{iq_y}{R_y}\Gamma^{\hat y}F+\frac{iq_z}{R_z}\Gamma^{\hat z}F
=0
\eeq 
It has normalized solutions  
\beq
F_{q_y, q_z}=\frac{\kappa^3\rho^3e^{-Q(\rho-\rho_0)}}{2\sqrt{2}\pi 
R_yR_z(q_z^2R_y^2+q_y^2R_z^2)^{1/4}}
\l(
\begin{array}{r}
\sqrt{q_z^2R_y^2+q_y^2R_z^2}\\q_z R_y+ iq_y R_z
\end{array}\r)
\eeq
where
\beq
\label{Q}
Q=\sqrt{\frac{q_y^2}{R_y^2}+\frac{q_z^2}{R_z^2}} 
\eeq
The Dirac mass $M_D$ which is the mixing between the modes 
$\Psi_{q_{yL},q_{zL}}$ and $\Psi_{q_{yR}, q_{zR}}$ is given by the 
integral over the extra dimensions of the overlap of the profiles 
of left and right handed components 
with the profile of the Higgs field (see (\ref{mq})). In the case of the sharp profile (\ref{sharp})
of the Higgs 
field $M_D$ is proportional to
\beq
\label{hie}
M_D\sim \exp\left\{-\l(\sqrt{\frac{q_{yL}^2}{R_y^2}+\frac{q_{zL}^2}{R_z^2}}-
\sqrt{\frac{q_{yR}^2}{R_y^2}+\frac{q_{zR}^2}{R_z^2}}\r)(\rho_h-\rho_0)\right\}
\eeq
(compare with (\ref{m2})).
If the scales $R_y, R_z$ and $(\rho_h-\rho_0)$ are arranged as 
$R_y\sim (several) R_z$, \  $(\rho_h-\rho_0)\sim (several) R_z$, 
the  masses of particles with different $q_{zL}, q_{zR}$  go as different 
powers of a small parameter
\beq
\label{eps}
M_D\sim \epsilon^{(q_{zL}-q_{zR})}
\eeq
where $\epsilon$ is 
\beq
\label{eps1}
\epsilon=\exp\left\{-\frac{\rho_h-\rho_0}{R_z}\right\}
\eeq 
The rotation symmetry $U(1)_z$ can be identified with the ``horizontal''
symmetry \cite{generation} which is introduced in some approaches to the 
fermion mass hierarchy and enables to distinguish between fermions 
from different generations.  The mass hierarchy of Eq. (\ref{eps}) 
is similar to the formula 
for the fermion masses derived in the models with the horizontal 
symmetry. In order to get a realistic pattern of masses of the standard model 
fermions the parameter $\epsilon$ must have numerical value 
\beq
\epsilon\approx 0.049
\eeq
which means that, if the thickness of the brane 
$\rho_0$ is negligibly small,  the relation between the radius of the 
Higgs orbit $r_h$ and the size of the circle $S^1$ parameterized by the 
coordinate $z$ must be
\beq
\rho_h\approx 3 R_z
\eeq
From (\ref{hie}) one can see that the 
masses of fermions with different charges $q_{yL}, q_{yR}$ are 
quasidegenerate.

\section{A model with $SU(2)\times U(1)$ symmetry.}
\label{monop}

Up to now we have considered models in which fermions carry only 
$U(1)$ charges. If we want to include $SU(2)\sim SO(3)$ 
as a symmetry group of rotations around 
the brane, a natural generalization of the model of the previous sections 
would be to consider a brane embedded in an eight-dimensional space
\beq
\label{metric3}
ds^2=e^{\nu(r)}\eta_{\mu\nu}dx^\mu dx^\nu+e^{\mu(r)}dy^2
+e^{\lambda(r)}\l(dr^2+r^2(d\theta^2+\sin^2\theta d\phi^2)\r)
\eeq 
that is, to add two more extra dimensions $(\theta, \phi)$ with the geometry of
two-dimensional sphere.

In the previous sections 
we have systematically abandoned the discussion of behavior of the fermion 
zero modes in the core of the brane. At the same time, if the brane is 
considered as a topological defect in a higher-dimensional space-time, the 
analysis of behavior of fermion zero modes in the core of defect is 
important, because the requirement of regularity of the fermion profiles 
in the core can impose restrictions on the number of normalizable 
zero modes bound to the brane \cite{jackiw}. 
In this section we will extend the analysis of fermion zero modes to the 
brane core. The topology of the core of the brane embedded into 
eight-dimensional space-time (\ref{metric3}) depends on the behavior of 
warp factors $\nu(r),\lambda(r), \mu(r)$ in the limit $r\rightarrow 0$ 
\cite{neronov1}. We consider the situation when 
\beq
\label{r->0}
\nu(r),\mu(r),\lambda(r)\rightarrow 0, \ \ r\rightarrow 0
\eeq  
so that the topology of the brane core is 
$R^4\times S^1\times D^3$, where $D^3$ is the three-dimensional disk 
$(r, \theta,\phi),\ 0\le r\le r_0$.  
  
The solutions of the eight-dimensional Dirac equation 
can be expanded over the eigenstates of the total angular momentum
\beq
\label{angular}
\vec L=\vec r\times\vec p+\frac{1}{2}\vec \sigma
\eeq
Here $\vec r\times \vec p$ is the usual angular momentum in three-dimensional 
space $(r, \theta, \phi)$ 
and $\frac{1}{2}\vec \sigma$ is the spin operator,  
($\sigma^i$ are the Pauli matrices).
If we relate $SU(2)$ symmetry of space-time (\ref{metric3}) 
to the $SU(2)_L$ gauge group of the 
standard model, we face immediately  the following difficulty. It is known,
that the right-handed fermions of the standard model 
are singlet representations  
of $SU(2)_L$. But the lowest possible value 
of the angular momentum (\ref{angular}) is 
\beq
l_{min}=1/2.
\eeq 
The state with the lowest angular momentum
is a doublet representation of $SU(2)$. 

This difficulty can be resolved if we suppose that the space-time
(\ref{metric3}) is obtained in result of Freund-Rubin compactification
\cite{freund} of higher-dimensional theory. For example, 
the space-time with geometry
\beq
\label{adss2}
ds^2=\frac{1}{\kappa^2 r^{2\kappa{\cal R}}}
\l(\eta_{\mu\nu}dx^\mu dx^\nu+dy^2\r)+
\frac{{\cal R}^2}{r^2}\l(dr^2+r^2(d\theta^2+\sin^2\theta d\phi^2)\r)
\eeq
of direct product $AdS_6\times S^2$ of anti-deSitter space with a two-sphere
of the radius ${\cal R}$ 
can be obtained as a solution of eight-dimensional 
Einstein-Maxwell equations 
if we  introduce a vector field
${\cal A}_M$ in a topologically nontrivial monopole-like 
configuration
\beq
\label{monopole}
{\cal A}_\phi=m(1-\cos\theta)
\eeq
in the eight-dimensional bulk. The metric (\ref{adss2}) can describe the 
space-time  geometry outside the brane, $r_0<r$, or asymptotically when 
$r\rightarrow\infty$,  while in the core of 
the brane, $0\le r\le r_0$,  
the metric is a solution of the Einstein-Maxwell equations coupled to other 
matter fields and the asymptotic behavior of the metric in the 
$r\rightarrow 0$ limit is given by (\ref{r->0}).

If the higher-dimensional spin-1/2 field $\Psi$ has nonzero charge $e$ with 
respect to ${\cal A}_M$, the conserved total  angular momentum is 
\beq 
\label{total}
\vec J=\vec r\times\vec p+\frac{1}{2}\vec \sigma-em\frac{\vec r}{r}
\eeq
rather then (\ref{angular}). The lowest 
possible value of the angular momentum is 
\beq
\label{lowest}
j_{min}=|em|-1/2
\eeq
and if $|em|=1/2$ the state 
with the lowest angular momentum is the $SU(2)$ singlet. The properties of 
the fermion zero modes in the model with monopole-like configuration 
(\ref{monopole}) of the  vector field ${\cal A}_M$ on the two-sphere $S^2$
were first considered in \cite{salam}.

In fact, the existence  of a fundamental vector field ${\cal A}_M$ in the 
eight-dimensional bulk is not a necessary  condition for the existence of 
$SU(2)$ singlets in the spectrum of fermionic modes. If we consider 
a little bit more complicated Ansatz for the background metric 
in the eight-dimensional space-time
\beq
\label{KK}
ds^2=e^{\nu(r)}\eta_{\mu\nu}dx^\mu dx^\nu+
e^{\mu(r)}\l(dy+{\cal A}_\phi d\phi\r)^2+
e^{\lambda(r)}\l(dr^2+r^2(d\theta^2+\sin^2\theta d\phi^2)\r)
\eeq
where ${\cal A}_\phi$ is given by (\ref{monopole}), the conserved angular 
momentum is again given by (\ref{total}). The only difference between 
the above two ways of introducing monopole-like configuration  
(\ref{monopole}) in the higher-dimensional theory is that in the Kaluza-Klein 
case (\ref{KK}) the Dirac field has a
nominimal coupling to the field ${\cal A}_M$ \cite{calmet}:
\beq
\label{nonmin}
L_{int}\sim \frac{m}{4r^2}e^{\mu/2-\lambda}\overline\Psi\Gamma^{\hat \theta}
\Gamma^{\hat \phi}\Gamma^{\hat
y}\Psi
\eeq
and the charge $e$ is expressed 
through the rotation momentum in $y$ direction
\beq
\label{qe}
e=\frac{q_y}{R_y}
\eeq
In both cases (in the space-time (\ref{metric3}) with the fundamental 
vector field ${\cal A}_M$  or in the space-time (\ref{KK}) with 
Kaluza-Klein vector field ${\cal A}_M$) the Dirac equation  for the fermion 
zero modes 
($\Gamma^{\hat\mu}\Psi_{,\mu}=0$) reduces to
\begin{eqnarray}
\label{dirac2}
e^{-\lambda/2}\left[\Gamma^{\hat r}\left(\Psi'+
\left(\nu'+\frac{\mu'}{4}+\frac{\lambda'}{2}+\frac{1}
{r}\right)\Psi\right)+
\frac{1}{r}\Gamma^{\hat\theta}\l(\Psi_{,\theta}+\frac{1}{2}\cot\theta\Psi\r)
\right.\nn\\ \left.
+\frac{1}{r\sin\theta}\Gamma^{\hat\phi}\l(\Psi_{,\phi}-ie{\cal A}_\phi
\Psi\r)\right]+e^{-\mu/2}\Gamma^{\hat y}\Psi_{,y}+k\frac{m}{4r^2}
e^{\mu/2-\lambda}\Gamma^{\hat \theta}\Gamma^{\hat \phi}\Gamma^{\hat y}\Psi
=0
\end{eqnarray}
(we have introduced an arbitrary coefficient $k$ in front of the nonminimal 
coupling term in order to be able to analyze the cases with and without 
nonminimal coupling (\ref{nonmin}) simultaneously). 
It is convenient to write $\Psi$ in the form
\beq
\Psi_{q}=\frac{1}{r}\exp\left\{-\nu-\frac{\mu}{4}-\frac{\lambda}{2}\right\}
\exp\left\{\frac{iq_y y}{R_y}\right\}
e^{-i\Gamma^{\hat y}}F(r, \theta, \phi)
\eeq
where $F$ is a four-component spinor.
From (\ref{dirac2}) we find that $F$ is a zero-energy solution of the  
four-dimensional Dirac equation
\beq
\label{zero}
HF=0
\eeq
with the Hamiltonian
\beq
\label{hamilton}
H=\vec\alpha(\vec p-e\vec A)+\beta \frac{q_y}{R_y}e^{(\lambda-\mu)/2}
-k\frac{m}{4r^2}e^{(\mu-\lambda)/2}\beta
\frac{(\vec\sigma\vec r)}{r^3}
\eeq
Here the matrices $\vec \alpha$ and $\beta$ are 
\beq
\vec\alpha=\l(\begin{array}{cc}
0&\vec\sigma\\
\vec\sigma &0
\end{array}\right), 
\beta=\l(\begin{array}{cc}
1&0\\
0&-1
\end{array}\right)
\eeq
The Hamiltonian (\ref{hamilton}) coincides with the Hamiltonian of Dirac 
particle in the central field
\beq
\label{vr}
V(r)=\frac{q_y}{R_y}e^{(\lambda-\mu)/2}
\eeq
superposed with the field of magnetic monopole (\ref{monopole}). 
The last term in the Hamiltonian (\ref{hamilton}) describes an extra magnetic 
moment of Dirac particle. The bound states of Dirac particles with an extra 
magnetic moment in the field of magnetic monopole were studied by Kazama
and Yang 
\cite{kazama}. The state with the lowest value (\ref{lowest}) of the total 
angular momentum  (\ref{total}) 
and its projection $J_3=m$ on some axis is described by the spinor
\beq
\label{fg}
F=\l(\begin{array}{c}
f(r)\\
ig(r)\end{array}\r)\otimes \eta_{|em|, j, m}(\theta, \phi)
\eeq
where $\eta_{|em|, j, m}$ is the two-component spinor 
\beq
\label{eta}
\eta_{(eg), j,m}=\l(\begin{array}{c} 
-\sqrt{\frac{j-m+1}{2j+2}}Y_{(eg),\ j+1/2,\  m-1/2}\\
\sqrt{\frac{j+m+1}{2j+2}}Y_{(eg),\ j+1/2,\  m+1/2}
\end{array}
\r)
\eeq
and $Y_{q, j, m}(\theta, \phi)$ are the monopole harmonics \cite{wu}.
Substituting (\ref{fg}) into (\ref{zero}) we find that functions 
$f(r), g(r)$ satisfy equations
\beqa
f'&=&\l(\frac{q_y}{R_y}e^{(\lambda-\mu)/2}-k\frac{m}{4r^2}e^{(\mu-\lambda)/2}
\r)
g\nn\\
g'&=&\l(\frac{q_y}{R_y}e^{(\lambda-\mu)/2}-k\frac{m}{4r^2}e^{(\mu-\lambda)/2}
\r)f 
\eeqa
We are interested in the bound state solution 
\beq
\label{solution}
f=-g=C\exp\left\{-\int dr\l(\frac{q_y}{R_y}e^{(\lambda-\mu)/2}-
k\frac{m}{4r^2}e^{(\mu-\lambda)/2}\r)\right\}
\eeq
($C$ is a normalization constant). Suppose that at large $r$ the bulk metric 
approaches  the metric (\ref{adss2}) of $AdS_6\times S^2$ space.
Then the  asymptotic behavior of (\ref{solution}) is
\beq
f=-g\sim\exp\left\{-\frac{q_yr^{\kappa {\cal R}}}{R_y}\right\}, 
\  \ r\rightarrow\infty
\eeq
This profile is essentially the same as  (\ref{profile}) if we relate the 
coordinates  $r$ and $\rho$
\beq
\rho=r^{\kappa{\cal R}}
\eeq
In the core of the brane the functions $\mu(r),\lambda(r)$ behave as 
in (\ref{r->0}) and the profile (\ref{solution}) is
\beq
f=-g\sim
\exp\left\{-k\frac{m}{4r}\right\},\ \ r\rightarrow 0
\eeq
If $q_y>0$, $f$ and $g$ vanish both when $r\rightarrow\infty$ and $r\rightarrow 0$ and the profile (\ref{solution}) corresponds to 
a fermion state localized on the brane. 

Let us suppose that the brane thickness $r_0$ is quite small, so that the 
region $r>r_0$ gives the main contribution in the normalization 
(\ref{norm}) of the fermion modes.
If the metric outside the brane is given by (\ref{adss2}), the normalized 
solution of Dirac equation which describes an $SU(2)$ singlet state is
given by (see (\ref{fg}), (\ref{solution}))
\beq
\label{singlet}
F_{q,0,0}=\frac{\sqrt{|q_y|}}{\sqrt{\pi}R_y}e^{-|q_y|(\rho-\rho_0)/R_y}
\l(\begin{array}{c}
1\\
i\end{array}\r)\otimes \eta_{1/2, 0, 0}(\theta, \phi)
\eeq
while the $SU(2)$ doublet state has the profile 
\beq
\label{duplet}
F_{q,1/2,m}=\frac{\sqrt{|q|}}{\sqrt{\pi} R_y}e^{-|q_y|(\rho-\rho_0)/R_y}
\l(\begin{array}{c}
1\\
i\end{array}\r)\otimes \eta_{1, 1/2, m}(\theta,\phi)
\eeq
If the monopole field (\ref{monopole}) appears as a 
Kaluza-Klein field (\ref{KK}), then the charges $q_y$ and  $e$ are related 
through (\ref{qe})
and instead of the whole tower of singlet and doublet states (\ref{singlet}), 
(\ref{duplet}) with different $q_y=1, 2,...$, we get just one 
singlet with $q_y=1$ and the only doublet with $q_y=2$.

The effective four-dimensional Dirac masses of localized fermions can be 
calculated by the same procedure as in the previous section. 
The conserved angular momentum for the Higgs scalar in the field of magnetic 
monopole (\ref{monopole}) is given by (\ref{total}) without the spin 
operator $\frac{1}{2}\vec \sigma$. If the charge of the Higgs field is   
 $e_h=1/(2m)$ the lowest possible value of the angular momentum is 
$j_h=1/2$. Therefore the higher-dimensional profile of 
the Higgs field is 
\beq
\label{higgs2}
H_{p, j=1/2, m=1/2}=H_{p}(r)e^{ipy/R_y}Y_{1/2, 1/2, 1/2}(\theta, \phi)
\eeq 
where $Y_{1/2, 1/2, 1/2}$ is the corresponding monopole harmonic \cite{wu}. 
In the same way as in (\ref{MD}) the Dirac mass of a fermion 
is given by the integral over the extra dimensions of the overlap of the 
singlet (\ref{singlet}) and doublet (\ref{duplet}) profiles
with the profile of the Higgs field (\ref{higgs2})  
\beq
\label{last}
M_D=f\int drdyd\theta d\phi\  e^{(\lambda-\nu)/2}H_{p,1/2, 1/2}
\overline F_{q_R, 0,0}F_{q_L,1/2,m}
\eeq  
Integrating over the angles $(\theta, \phi)$ we find that the mixing of 
the doublet state with projection of angular momentum 
$m=-1/2$ with the singlet vanishes. 
The integral (\ref{last}) for the mixing of $m=1/2$ component of 
the doublet with the singlet reduces to (\ref{m}). 
The mass $M_D$ is expressed through the charges $q_L$ and $q_R$ 
of doublet (left) and singlet (right) components in the same way as in the 
previous Section. Depending on the profile of the Higgs field $H_p$ 
we can get either hierarchical (\ref{m2}) or quasidegerate (\ref{m1}) 
pattern of fermion masses.

It is straightforward to generalize the above model to the case when the 
brane is embedded in the nine-dimensional 
space-time with asymptotic $AdS_7\times S^2$ 
geometry. In this case the localized fermion modes are charged with respect 
to the $SU(2)\times U(1)\times U(1)$ symmetry group. One of the $U(1)$ 
factors can be related to the $U(1)_Y$ gauge group of the standard model, 
while the other will correspond to the horizontal symmetry $U(1)_G$.
 There are $SU(2)$ singlets and doublets 
in the spectrum of localized fermions. After the symmetry breaking the upper
components of doublets become massive. The masses of these states 
are proportional to different powers of a small parameter $\epsilon$
(\ref{eps}), (\ref{eps1}) and have hierarchical structure. 
Correspondingly, these states can be identified 
with the charged leptons of the standard model. The lower components of the 
$SU(2)$ doublets remain massless, since they are not mixed to the states 
localized on the brane. These states should be identified with neutrinos. 

In our model the higher-dimensional profiles of the fermions are completely 
determined by their quantum numbers. In order to include quarks in the model 
we must equip the fermions with $SU(3)$ quantum numbers. We leave the 
discussion of relation between quark masses and quantum numbers for the 
future work.

\section{conclusion.}

In this paper we have pointed out that in the models with a brane universe embedded into 
a higher-dimensional space-time, the fermions localized on the brane 
always possess quantum numbers which are their rotation momenta around the 
brane. We have considered 
the cases where the localized fermions carry $U(1)$ (Section \ref{ads6}), 
$U(1)\times U(1)$ (Section \ref{masses}) or $SU(2)\times U(1)$ 
(Section \ref{monop}) quantum numbers. 

The higher-dimensional profiles of the fermions depend on their quantum 
numbers (\ref{profile}),  and it turns out that the fermions which are neutral 
with respect to the group of rotations around the brane are not localized. 

The effective four-dimensional masses of localized fermions are proportional to
the overlaps of the profiles 
$\Psi_L, \Psi_R$ of left
and right handed components of the fermion with the profile 
$H_p$ of (effective or fundamental)  Higgs field (\ref{MD}). If the Higgs 
field is sharply peaked at a distance $\rho_h$ from the center of the brane, 
like  in (\ref{sharp}), the the masses
of fermions with different quantum numbers go like different powers of 
a small parameter $\epsilon$, (\ref{eps}), (\ref{eps1}). Even if there is no
hierarchy in the length scales $\rho_h$ and  $R_z$ in (\ref{eps1}), the masses of localized fermions are arranged 
hierarchically.

\section{acknowledgment.}

I am grateful to A.Barvinsky, X.Calmet, V.Mukhanov, I.Sachs and S.Solodukhin
for useful discussions of the subject of the paper. This work was 
supported by the SFB 375 Grant of Deutsche Forschungsgemeinschaft.

\end{document}